# A New Proposed Technique to Improve Software Regression Testing Cost

Seifedine Kadry
*American University of the Middle East, Kuwait*
skadry@gmail.com

*Abstract*

*In this article, we describe the regression test process to test and verify the changes made on software. A developed technique use the automation test based on decision tree and test selection process in order to reduce the testing cost is given. The developed technique is applied to a practical case and the result show its improvement.*

*Keywords: software maintenance, software test, test automation, regression.*

## 1. Introduction

System maintenance is a general term required to keep a system running properly. The system could be computer system, mechanical system or others. The maintenance in this sense is related to the deterioration of the system due to its usage and age. This context of maintenance does not apply to software, where the deterioration due to the usage and age don't make sense. Conventionally, the maintenance of software is concerned with modifications related to software system. These modifications come from the user needs, error correction, improvement of performance, adapt to a changed environment, and optimization. Software development companies spend more time on maintenance of existing software than on development of new software, and according to earlier studies software maintenance accounts between 40-70% (figure 1) of its the total life-cycle costs.

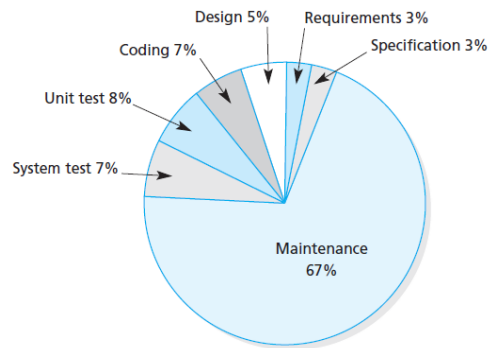

**Fig. 1: Costs of Software Development Stages [11]**

Due to the testing process of software, the maintenance phase is undoubtedly the most costly and crucial phase in the software development life cycle. In this chapter we study in details two well known testing techniques: Regression Test Selection and Automation test, in addition to the development of a new technique to ameliorate the cost-effectiveness of the regression testing. The proposed technique is applied to an insurance system in the SNA-Soft Company, the evaluation and a comparison with other techniques is given.

## 2. Maintenance Cost

As software systems age, it becomes more and more difficult to satisfy user requirements and to keep them 'up and running' without maintenance.

Maintenance is applicable to software developed using any software life cycle model (waterfall, spiral, etc). Maintenance must be performed in order to:

- Interface with other systems
- Correct faults
- Migrate legacy software
- Implement enhancements
- Adapt programs so that different hardware, software, system features, and telecommunications facilities can be used
- Improve the design
- Retire software

Some of the technical and non-technical factors affecting software maintenance costs, as follows:

- Team stability
- Application type
- Program age and structure
- Software novelty
- Stressful nature of work
- Software maintenance staff availability
- Software life span
- Staff skills
- Hardware characteristics

*For instance, in the United States, 2% of the GNP (Gross National Product) is spent on software maintenance and in UK; about $1.5 million annually are spent on software maintenance.*

## 3. Software Maintenance Types

E.B. Swanson initially identified three categories of maintenance: corrective, adaptive, and perfective. Later on, one additional category is added by Lientz and Swanson [1]: Preventive maintenance. The usage percentage of these categories is shown in figure 2.

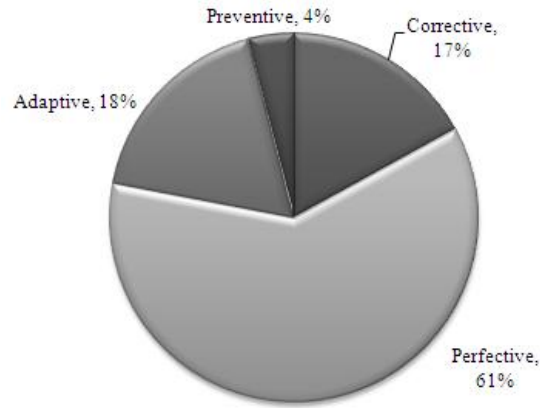

**Fig. 2: Maintenance Type's Percentage [12]**

Corrective maintenance is any maintenance activity to fixes the bugs that has occurred or after delivery [2], while adaptive maintenance keeps the software program working in a changed or changing environment such as the hardware or the operating system [3]. Perfective maintenance is every change to software to increase its performance or to enhance its user interface and its maintainability [4, 14]. The goal of preventive maintenance is to keep the software functioning correctly at any time.

## 4. Testing and Verification

After any modifications or maintenance, the verification step is very critical. Verification is the general term for techniques that aim to assure that software fully satisfies all the expected requirements without any "bugs". Testing is a widely used technique for verification, but note that testing is just one technique amongst several others. Currently the dominant technique used for verification is testing. And testing typically consumes an enormous proportion (sometimes as much as 50%) of the effort of developing a system. Microsoft employs teams of programmers (who write programs) and completely separate teams of testers (who test them). At Microsoft there are as many people involved in testing as there are in programming. Arguably, verification is a major problem and we need good techniques to tackle it. Often, towards the end of a project, the difficult decision has to be made between continuing the testing or delivering the software to the customers.

Code can be tested at many different levels with different techniques: do individual statements execute according to specification, do procedures provide expected output for given input, and does the program as a whole perform in a particular way? Within this are many issues to be borne in mind. For example, it is possible to execute each statement without touching upon certain conditions. However, test cases should try to take account of all possible conditions and combinations of conditions, with special emphasis on boundary conditions and values where behavior is often erroneous. In the following sections, we discussed some of the techniques used in software testing.

*Black Box*

Knowing that exhaustive testing is infeasible, the black box approach to testing is to devise sample data that is representative of all possible data. We then run the program, input the data and see what happens. This type of testing is termed black box testing because no knowledge of the workings of the program is used as part of the testing – we only consider inputs and outputs. The program is thought of as being enclosed within a black box. Black box testing is also known as functional testing because it uses only knowledge of the function of the program (not how it works). Ideally, testing proceeds by writing down the test data and the expected outcome of the test before testing takes place. This is called a test specification or schedule. Then you run the program, input the data and examine the outputs for discrepancies between the predicted outcome and the actual outcome. Test data should also check whether exceptions are handled by the program in accordance with its specification.

*White Box*

This form of testing makes use of knowledge of how the program works – the structure of the program – as the basis for devising test data. In white box testing every statement in the program is executed at some time during the testing. This is equivalent to ensuring that every path (every sequence of instructions) through the program is executed at some time during testing. This includes null paths, so an "if statement" without an else has two paths and every loop has two paths. Testing should also include any exception handling carried out by the program.

*Code Debugging*

Some debuggers allow the user to step through a program, executing just one instruction at a time. Each time you execute one instruction you can see which path of execution has been taken. You can also see (or watch) the values of variables. It is rather like an automated structured walkthrough. In this form of testing, you concentrate on the variables and closely check their values as they are changed by the program to verify that they have been changed correctly. A debugger is usually used for debugging (locating a bug); here it is used for testing (establishing the existence of a bug).

*Test Profiler*

In a large system or program it can be difficult to ensure that the test data is adequate. One way to try to test whether it does indeed cause all statements to be executed is to use a profiler. A profiler is a software package that monitors the testing by inserting probes into the software under test. When testing takes place, the profiler can expose which pieces of the code are not executed and therefore reveal the weakness in the data. Another approach to investigating the test data is called mutation testing. In this technique, artificial bugs are inserted into the program. An example would be to change a + into a –. The test is run and if the bugs are not revealed, then the test data is obviously inadequate. The test data is modified until the artificial bugs are exposed.

*Beta Testing*

In beta testing, a preliminary version of a software product is released to a selected market, the customer or client, knowing that it has bugs. Users are asked to report on faults so that the product can be improved for its proper release date. Beta testing gets its name from the second letter of the Greek alphabet. Its name therefore conveys the idea that it is the second major act of testing, following on after testing within the developing organization. Once Beta testing is complete and the bugs are fixed, the software is released.

*Test Automation*

It is good practice to automate testing so that tests can be reapplied at the touch of a button. This is extra work in the beginning but often saves time overall. The following reasons lead the software companies to apply a test automation process on software:

*Manual mistakes* – may occur due to human interference. In this direction, the execution of test automation reduces the wrong entering information by the tester.

*Parallel execution* – the computer is multitask, it can run many tests in parallel hence the process can be done faster with less resources.

*No Interruption* – there is no need to interrupt the execution of automated test, contrary to manually executed cycle.

*Easy Result Analysis* – the analysis result phase comes after the execution of automated test. Also, this phase can be automated based on previous test analysis result, which is easier to do it manually.

Although the automation process seems to have more benefits than disadvantages, it is always good to be aware of the problems that may be encountered. The anxiety of having everything automated is higher because the tester wants to feel comfortable by running all the tests several times, avoiding bug insertions when the code changes [5]. Some tests, even during reruns, may never find a bug and have no significant prospect of doing so. Hence automating everything may not be the best decision. To avoid these kinds of mistakes and have a cost-effective automation process, there are some tradeoffs that need to be well-understood before deciding whether a particular test should be automated.

To improve the automation test, the viability method (ATVM) is developed [5]. Some questions were proposed whose answers will be analyzed and judged properly to decide if a test should be automate it or not in order to improve its cost-effectiveness, as illustrated in Table 1.

**Table1: indicator Questions for Test Automation**

| Identifier | Topics | Related Questions |
|---|---|---|
| 1 | Frequency | How many efforts is this test supposed to be executed? |
| 2 | Reuse | Can this test or parts of it be reused in other tests? |
| 3 | Relevance | How would you describe the importance of this test case? |
| 4 | Automation effort | Does this test take a lot of effort to be deployed? |
| 5 | Resources | How many members of your team should be allocated or how expensive is the equipment needed during this test's manual execution? |
| 6 | Manual Complexity | Is this test difficult to be executed manually? Does it have any embedded confidential information? |
| 7 | Automation Tool | How would you describe the reliability of the automation tool to be used? |
| 8 | Porting | How portable is this test? |
| 9 | Execution effort | Does this require a lot of effort to be executed manually? |

The Decision Tree (figure 5) Learning Algorithm is based on the answer to the above questions (Table 1) of 500 automated test cases. The building process and verification of the decision tree is automatic.

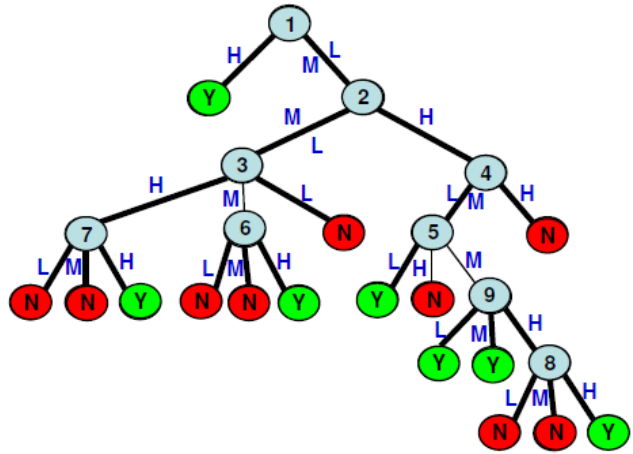

**Fig. 5: Decision Tree**

The tree represents the ways that can be followed by answering the questions presented in Table 1. The numbers in the grey circles are the Identifiers for the questions that they are related to. The 'Y' and 'N' represent the end of the tree and mean 'Yes' or 'No', respectively. They are the indicators of cost-effectiveness.

All the proposed questions have a discrete number of answers: 'High', 'Medium' or 'Low', which are represented in the tree by the letters 'H', 'M' and 'L'. Note that, depending on the answer of each specific question, the tree takes you different ways. Going through the tree from the top to the end via the answered questions, an indicator of 'Yes' or 'No' will be obtained, showing if a test case is viable or not for automation. It is important to notice that not all the questions need to be answered for a test case. For example: if the answer given to the question number 2 is 'L', question number 4 will never be answered. Also, it is worth clarifying that there is no correct answer for these questions. However, the more you know about the tests being analyzed, the better the chances will be for success.

Example: customer registration in SNA-Soft system. In this example, we illustrate how to apply the above decision tree for a test case.

Starting with the first question on the tree, "How many times is this test supposed to be executed?", if this operation has few executions, the answer is 'Low'. This answer takes you to the right side of the tree, leading to the second question, "Can this test or parts of it be reused in other tests?" Let's suppose that the code used to automate this test has little chance of being reused. Thus, the answer to Question 2 is 'Medium'. Now the Decision Tree takes us to the left side of the tree, to the next question, "How would you describe the importance of this test case?" Making a transaction on a bank website is an important task to be tested, so the answer is 'High'. The left side is taken, which leads to the last question, "How would you describe the reliability of the automation tool to be used?" As the test has very high relevance, the tool to be used must be quite reliable to ensure that this test is in fact being well executed. Therefore, the answer to this question is 'High'. The summary of the results reached with this example is presented in Table 2.

**Table 2: Questions Result**

| Identifier | Questions | Answers |
|---|---|---|
| 1 | How many efforts is this test supposed to be executed? | Low |
| 2 | Can this test or parts of it be reused in other tests? | Medium |
| 3 | How would you describe the importance of this test case? | High |
| 7 | How would you describe the reliability of the automation tool to be used? | High |

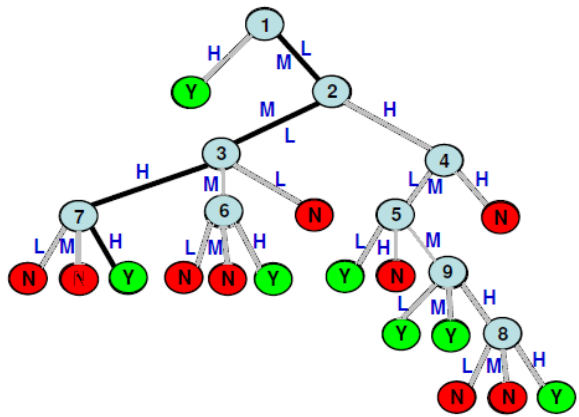

**Fig. 6: Decision Tree of the Questions Result**

By answering the questions 1, 2, 3 and 7, following the decision tree (the black lines, figure 67), the user would have a positive response, which would mark this test as a good candidate for automation. Note that it is not necessary to answer all the questions. Depending on the answers that are given, the tree can conduct the user to answer only some of the questions.

*Regression Testing*

Regression testing is defined as "the process of retesting the modified parts of the software and ensuring that no new errors have been introduced into previously tested code". Let P be a program, let P′ be a modified version of P, and let T be a test suite for P. Regression testing consists of reusing T on P′, and determining where the new test cases are needed to effectively test code or functionality added to or changed in producing P′. There is various regression testing techniques [6]: Retest all, Regression Test Selection, Test Case Prioritization and Hybrid Approach (Figure 7).

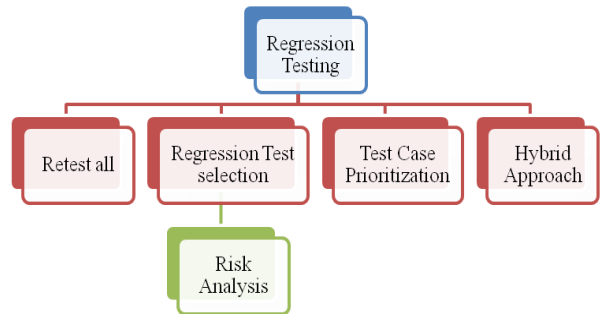

**Fig. 7: Regression Testing Techniques**

*Retest All*

Retest all method is one of the conventional methods for regression testing in which all the tests in the existing test suite are re-ran. So the retest all technique is very expensive as compared to techniques which will be discussed further as regression test suites are costly to execute in full as it require more time and budget.

*Test Case Prioritization*

This technique of regression testing prioritize the test cases so as to increase a test suite's rate of fault detection that is how quickly a test suite detects faults in the modified program to increase reliability. This is of two types: (1) General prioritization which attempts to select an order of the test case that will be effective on average subsequent versions of software. (2)Version specific prioritization which is concerned with particular version of the software.

*Hybrid Approach*

The fourth regression technique is the Hybrid Approach of both Regression Test Selection and Test Case Prioritization. There are number of researchers working on this approach and they have proposed many algorithms for it.

*Regression Test Selection*

Due to expensive nature of "retest all" technique, Regression Test Selection (RTS) is performed. In this technique instead of rerunning the whole test suite we select a part of test suite to rerun if the cost of selecting a part of test suite is less than the cost of running the tests that RTS allows us to omit. RTS divides the existing test suite into (1) Reusable test cases; (2) Re-testable test cases; (3) Obsolete test cases. In addition to this classification RTS may create new test cases that test the program for areas which are not covered by the existing test cases. RTS techniques are broadly classified into three categories.
  1) Coverage techniques: they take the test coverage criteria into account. They find coverable program parts that have been modified and select test cases that work on these parts.
  2) Minimization techniques: they are similar to coverage techniques except that they select minimum set of test cases.
  3) Safe techniques: they do not focus on criteria of coverage; in contrast they select all those test cases that produce different output with a modified program as compared to its original version. Rothermel [7] identified the various categories in which Regression Test Selection Technique can be evaluated and compared. These categories are: (a) Inclusiveness; (b) Precision; (c) Efficiency; (d) Generality.
      a) Inclusiveness is the measure of extent up to which a technique chooses the test cases which will cause the changed program to produce different output than the original program, resulting in exposure of faults due to modifications.
      b) Precision is the measure of ability of technique to prevent choosing test cases that will not make the changed program to produce different output than the original program.
      c) Efficiency measures the practicality (computational cost) of a technique.
      d) Generality is the measure of ability of a technique to handle complex modifications, realistic language constructs and realistic testing applications.

*Regression test selection based on risk analysis*

This section is based on the study of Chen and Probert [13]. Risk is anything that threatens the successful achievement of a project's goals. Specifically, a risk is an event that has some probability of happening, and that if it occurs, will result in some loss. The tester's job is to reveal high-priority problems in the product. Traditional testers have always used risk-based testing, but in an ad hoc fashion based on their personal judgment. Using risk metrics to quantitatively measure the quality of a test suite seems perfectly reasonable and is our approach.

Amland [8] presented a simple risk model with only two elements of Risk Exposure. We use this model in our research. It takes into account both:

1. The probability of a fault being present. Myers [9] reports that as the number of detected errors increases, the probability that more undetected errors exist also increases. If one component has defects that are detected by full testing, it is very likely that we can find more defects in this component by regression testing (the more defects detected, the more defects we can expect). Thus, components with detected defects should be covered more carefully by regression tests.
2. The cost (consequence or impact) of a fault in the corresponding function if it occurs in operation. It is a well-known, observed fact that most commercial software contains bugs at delivery time. Companies begin a project with the knowledge that they will choose, because of schedule pressures, to ship software that contains known bugs [10]. However, when the budget is limited, we strive to detect the most critical defects first.

The mathematical formula to calculate Risk Exposure is $RE(f) = P(f) \times C(f)$, where $RE(f)$ is the Risk Exposure of function f, $P(f)$ is the probability of a fault occurring in function f and $C(f)$ is the cost if a fault is executed in function f in operational mode. In this context, we calculate first the Risk Exposure for each test case ($RE(t)$). Then we choose test cases based on $RE(t)$. To illustrate this formula we present a real example based on our SNA-Soft product. The evaluation of risk exposure involves 4 steps:

**Step 1**. Assess the cost for each test case

Cost means the cost of the requirements attributes that this test case covers. Cost C is categorized on a one to five scale, where one is low and five is high. Two kinds of costs will be taken into consideration:
- The consequences of a fault as seen by the customer, that is, losing market share because of faults.
- The consequences of a fault as seen by the vendor, that is, high software maintenance cost because of faults.

Table 3 shows the costs for some test cases in our case study, i.e. SNA-Soft insurance software.

**Table 3: Cost of Test Cases**

| Test Case (ct) | Cost C(ct) | Description |
|---|---|---|
| 1000 | 5 | Update the balance after payment |
| 1010 | 5 | Write changes to the log file |
| 1020 | 4 | Add new product to the customer profile |
| 1030 | 3 | Send email to the customer of the modifications |
| … | … | … |

**Step 2** Derive severity probabilities for each test case

After running the full test suite, we can sum up the number of defects uncovered by each test case. Learning from real in-house testing, we find that the severity of defects (how important or serious the defect is) is very important for software quality. Considering this aspect, we modify the simple risk model that we mentioned previously. The probability element in the original risk model is changed to severity probability, which combines the average severity of defects with the probability. Based on multiplying the Number of Defects N by the Average Severity of Defects S (N×S), we can estimate the severity probability for each test case.

Severity probability falls into a zero to five scales, where zero is low and five is high. For the test cases without any defect, P (t) is equal to zero. For the rest of the test cases, P (t) for the top 20% of the test cases with the highest estimate N×S will be five, and P (t) for the bottom 20% of the test cases will be one. Table 4 displays P (t) for some of the test cases in our case study.

**Table 4: Severity Probability of Test Cases**

| Test Case (ct) | Number of Defects (N) | Average severity of Defects (S) | NxS | P(t) |
|---|---|---|---|---|
| 1000 | 1 | 2 | 2 | 2 |
| 1010 | 1 | 3 | 3 | 2 |
| 1020 | 2 | 3 | 6 | 4 |
| 1030 | 0 | 0 | 0 | 0 |
| … | … | … | | |

**Step 3** Calculate Risk Exposure for each test case

Combining Table 3 and Table 4, we can calculate Risk Exposure RE (t) for each test case as shown in Table 5. To improve or focus the results, we can also add weights to test cases that we need to give preference to, for example, we might like to choose more test cases for some corporate flag-ship features, such as database features.

**Table 5: Risk Exposure of Test Cases**

| Test Case (ct) | Cost C(ct) | Number of Defects (N) | Average severity of Defects (S) | NxS | P(t) | RE(t) |
|---|---|---|---|---|---|---|
| 1000 | 5 | 1 | 2 | 2 | 2 | 10 |
| 1010 | 5 | 1 | 3 | 3 | 2 | 10 |
| 1020 | 4 | 2 | 3 | 6 | 4 | 16 |
| 1030 | 3 | 0 | 0 | 0 | 0 | 0 |
| … | … | … | … | | | |

**Step 4** Select Safety Tests

To get "good enough quality," our regression test suite should achieve some coverage target. This coverage target has to be set up depending on the available time and budget. Intuitively, we do not want to choose low-risk test cases, we choose test cases that have the highest value of RE (t).

## 5. The proposed technique

The main goal to use regression test selection based on risk analysis or automation test is to improve the cost-effectiveness of the test. Both of these techniques are expensive in resources sense. The new proposed technique is to combine these two techniques to obtain more improvement on the test cost-effectiveness. The flowchart of the proposed technique is given by (figure 8):

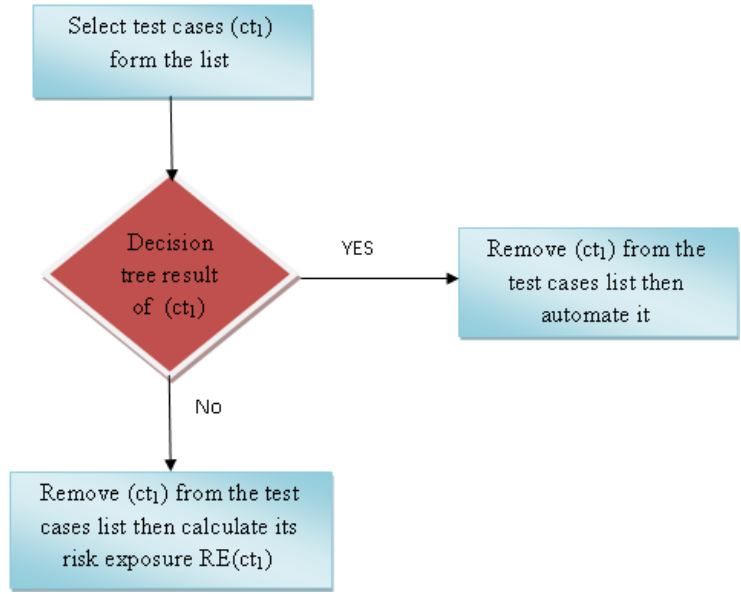

**Fig. 8: Proposed Technique Flowchart**

The process begin by selecting a test case, then to automate it or to calculate its risk exposure based on the result of the decision tree, then repeat the same process to finish all test cases.

## 6. Evaluation

To determine the influences on the cost-effectiveness of regression testing our evaluation methodology is based on three factors: Execution time complexity, number of errors detected, deployment time. Therefore the three techniques: Regression test selection based on risk analysis (TSRA), Test automation based on viability method (ATVM) and the proposed technique (PT) are evaluated based on these factors. The environment test is always the insurance software by SNA-Soft Company. This software is devoted to providing and managing all insurance products globally and it includes 20 modules. During the evaluation, we choose three different versions of the software. Each version is updates of the previous one, some errors are always expected. The threshold of TSRA was 70% of the total test cases with highest risk exposures. The following figure shows the execution time of the three methodologies. The average execution time of: TSRA is 13.6 hours, ATVM is 4.3 hours and PT is 2.6 hours. ATVM is more cost-effective of TSRA but PT is more cost-effective of both (figure 9).

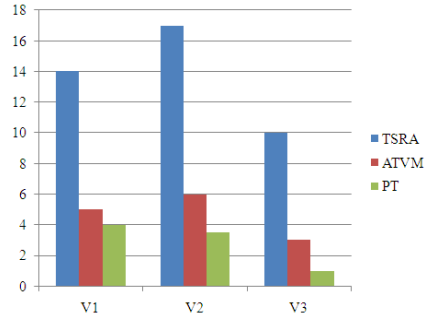

**Fig. 9: Execution Time Factor**

The difference in the results between different software versions was mainly caused by the bugs that were detected. When a bug was detected, we analyzed and tried to reproduce it by re-running the detecting test suite. These actions generally increased testing effort with all three methods. With TSRA, updating, calculating risk-exposures and selecting test suites required extra time. Figure 10 shows the number of errors detected with each method. Software version 3 did not contain any detected errors.

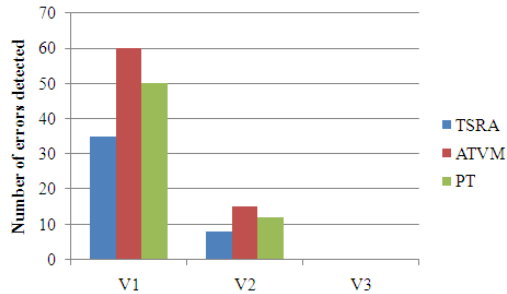

**Fig. 10: Number of Errors Factor**

The TSRA missed some errors because it was set to reduce the number of tests effectively; it covers only 70% of the overall test coverage. The number of undetected errors in the proposed technique (PT) is less than the number of undetected errors using TSRA because some of the test cases are automated and others ran based on the their risk exposure. ATVM detected all errors.

Figure 11 shows the deploying time (design and implementation) of the three methods. The most consuming time in the deploying phase is the design and the implementation of a script to automate the test cases. Figure 12 reveals that 60% of the overall effort consumed in deploying ATVM, 30% for the proposed technique and only 10 % for TSRA.

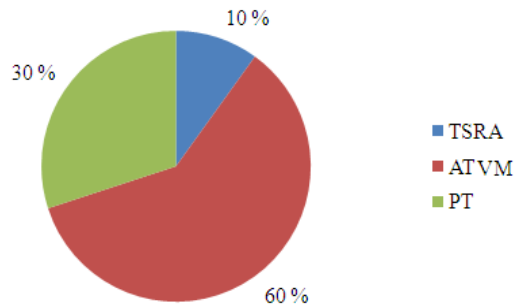

**Fig. 11: Deploying Time**

Based on the previous comparisons, the proposed technique could be utilized in larger scale as a method for improving the cost-effectiveness of regression testing. It's a compromise between TSRA and ATVM.

Congratulations! Your paper has been accepted for journal publication. Please follow the steps outlined below when submitting your final draft to the SERSC Press. These guidelines include complete descriptions of the fonts, spacing, and related information for producing your proceedings manuscripts. Please follow them and if you have any questions, direct them to the production editor in charge of your journal at the SERSC, sersc@sersc.org.

## 7. Conclusion

The objective of this chapter is to develop a new technique to improve the cost-effectiveness of the regression testing. An evaluation between the proposed technique and two well known techniques: Automated test using viability method and Test selection based on Risk Analysis is given. The evaluation is based on three factors: execution time, number of detected errors and the deploying time. We recommend using the proposed technique as a compromise technique between TSRA and ATVM.

## Authors

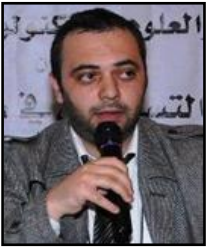
**Dr. Seifedine Kadry** is an associate professor at the American university of the Middle East, Faculty of Engineering. He got his Master Degree in Computer Science and Applied Math from AUF-EPFL-Inria, Lebanon in 2002. He received the Doctor degree from the Clermont Ferrand II University, France in 2007. His Research interests include software testing and security.